\documentclass[12pt]{article}       % onecolumn (second format)
\usepackage{authblk}
\usepackage{graphicx}
\usepackage{mathptmx}
\usepackage{amsmath}
\usepackage{color}
\usepackage[normalem]{ulem}
\newcommand{\be}{\begin{equation}}
\newcommand{\ee}{\end{equation}}
\newcommand{\bea}{\begin{eqnarray}}
\newcommand{\eea}{\end{eqnarray}}

\def\a{\alpha}

\def\g{\gamma}
\def\G{\Gamma}
\def\d{\delta}
\def\D{\Delta}
\def\e{\epsilon}

\def\l{\lambda}

\def\r{\rho}

\def\S{\Sigma}

\def\w{\omega}

\def\Q{\Psi}

\def\blk{{\mathbf k}}

\def\blp{{\mathbf p}}
\def\blq{{\mathbf q}}

\def\blx{{\mathbf x}}

\def\blP{{\mathbf P}}

\def\callL{\mbox{$\mathcal{L}$}}

\def\ra{\rightarrow}

\def\ket{\rangle}

\def\1op{\hat{\mathbbm{1}}}
\def\1{\mathbbm{1}}
\def\nn{\nonumber}

\newcommand{\nbar}{\overline{n}}
\newcommand{\kbar}{\overline{\blk}}

\newcommand{\tbar}{\overline{t}}

\newcommand{\rhom}{\underline{\rho}}
\newcommand{\hm}{\underline{h}}
\newcommand{\dm}{\underline{\d}}
\newcommand{\Sm}{\underline{\S}}
\newcommand{\Gm}{\underline{G}}
\newcommand{\Gtm}{\underline{\Gt}}
\newcommand{\Imm}{\underline{I}}
\newcommand{\Kmm}{\underline{\underline{K}}}
\newcommand{\Hmm}{\underline{\underline{H}}}
\newcommand{\Lmm}{\underline{\underline{L}}}
\newcommand{\Ltmm}{\underline{\underline{\Lt}}}
\newcommand{\Gmm}{\underline{\underline{\G}}}

\newcommand{\excop}{\hat{e}_{\gamma \blq}}
\newcommand{\excdop}{\hat{e}^\dagger_{\gamma \blq}}

\newcommand{\excxop}{\hat{e}_{x \blq}}
\newcommand{\excxdop}{\hat{e}^\dagger_{x \blq}}

\newcommand{\bosop}{\hat{a}_\alpha}
\newcommand{\bosdop}{\hat{a}^\dagger_\alpha}

\newcommand{\ecop}{\hat{c}_{\blk}}
\newcommand{\ecdop}{\hat{c}^\dagger_{\blk}}
\newcommand{\ecpqop}{\hat{c}_{\blk+\blq}}
\newcommand{\ecpqdop}{\hat{c}^\dagger_{\blk+\blq}}
\newcommand{\evop}{\hat{v}_{\blk}}
\newcommand{\evdop}{\hat{v}^\dagger_{\blk}}

\newcommand{\ekq}{\epsilon_{\blk\blq}}
\newcommand{\fkq}{f_{\blk\blq}}
\newcommand{\ekpq}{\epsilon_{\blk'\blq}}
\newcommand{\fkpq}{f_{\blk'\blq}}
\newcommand{\Lt}{\tilde{L}}
\newcommand{\Gt}{\tilde{G}}

\begin{document}

\title{An ab-initio approach to describe coherent and non-coherent exciton dynamics}

\author[1,3]{Davide Sangalli}
\author[1,3]{Enrico Perfetto}
\author[2,3]{Gianluca Stefanucci}
\author[1,3]{Andrea Marini}
%\authorrunning{Short form of author list} % if too long for running head

\affil[1]{CNR-ISM, Division of Ultrafast Processes in Materials (FLASHit),  Area della Ricerca di Roma 1, Monterotondo Scalo, Italy}
\affil[2]{Dipartimento di Fisica, Universit\`{a} di Roma Tor Vergata, and
INFN, Sezione di Roma Tor Vergata,
Via della Ricerca Scientifica 1, 00133 Roma, Italy}
\affil[3]{European Theoretical Spectroscopy Facility (ETSF)}

\setcounter{Maxaffil}{0}
\renewcommand\Affilfont{\itshape\small}

\maketitle

\begin{abstract}
The use of ultra-short laser pulses to pump and probe materials activates a wealth of
processes which involve the coherent and non coherent dynamics of interacting electrons out of equilibrium.
Non equilibrium (NEQ) many body perturbation theory (MBPT) offers an equation of motion for the
density--matrix of the system which well describes both coherent and non coherent processes.
In the non correlated case there is a clear relation between
these two regimes and the matrix elements of the density--matrix.
The same is not true for the correlated case, where the potential binding of electrons and holes in excitonic states need to be considered.
In the present work we discuss how NEQ-MBPT can be used to describe the dynamics of both coherent
and non-coherent excitons in the low density regime. The approach presented is well suited for an ab initio implementation.

\end{abstract}

\section*{Introduction}
%=====================================================

The concept of coherent states~\cite{Glauber1963a,Glauber1963b}
has been developed in the field of quantum optics, where photons states are considered.
It is not very familiar in the community of condensed matter and material science where the focus is on the
description of the electronic system in terms of non coherent Fock states.
One of the reasons is that coherences die very quickly in many-body electronic systems due to
the strong electronic interaction and cannot be easily observed. Moreover the concept of coherent states is 
associated to bosons: many bosons can occupy the same quantum states, bringing to the
manifestation of quantum coherence at the macroscopic or classical level. 
In very rare situations coherent states can be created by pairing fermions in ``quasi-bosons''.
A well known example is the superconducting state where the effective interaction brings to the formation and
condensation of Cooper pairs, thus of a coherent state which is stable, at least at low temperatures~\cite{Fetter-book}.
Another example is the excitonic insulator proposed by Kohn, where electron-hole ($eh$) pairs spontaneously bind into
excitons and then condensate~\cite{Sherrington1968}.
Coherent states for strongly interacting fermions are however an exception in the stationary regime.
In pump and probe experiments instead, materials are explored on a short time-scales such that coherences
are routinely observed.
The emergence and interplay between coherent and non-coherent dynamics is an example of the richness of
phenomena which can be observed in the non-equilibrium (NEQ) regime.
Indeed the notion of coherent and non coherent dynamics is
familiar to scientists working to the modeling of materials out of equilibrium~\cite{kb-book,hj-book}.
However, with the exception of few recent works~\cite{Ruggenthaler2014,Flick2015},
how the \emph{coherent dynamics} is related to the concept of \emph{coherent states} in quantum
optics is not well explored.

The interaction of the ultra-short (optical) pump pulse with a material leads to the formation
of NEQ states which are well described in terms of excitons. The existence of the two regimes,
i.e. coherent and non coherent, naturally brings to the definition of coherent and non coherent
excitons~\cite{Koch2006,Sipe2009b}. 
We will thus try to make a connection between the concept of \emph{coherent dynamics} from NEQ and the
concept of \emph{coherent states} from quantum optics for the case of the exciton~\cite{Wannier1937}.
The exciton is an interesting case because it is a \emph{composite boson}.
The operator defining the creation of an exciton, $\excdop$, can be written as a linear combination of $eh$ pairs.
This enables to use standard many-body perturbation theory (MBPT), formulated in Fock space,
since both  $\langle\excop\rangle$ and $\langle\excdop\excop\rangle$, can be expressed in terms of Green's functions.
As we will see the two expectation values are related to coherent and non-coherent states.
To give an intuitive picture a coherent excitons is related to the oscillations of the polarization of the system
resonant with the excitonic energies~\cite{Attaccalite2011}. It can be measured in absorption experiments. A non-coherent exciton instead is a
quasi-stationary state, which is a good approximation to neutral eigenstates of the many-body hamiltonian.
Their signature can be seen, for example, in photo-emission experiments~\cite{Perfetto2016,Rustagi2018,Steinhoff2017}.
The definition of $\excdop$ is a result of the equations of linear response theory, where everything can
be defined in terms of equilibrium quantities. To avoid that such definition will change in time in the NEQ regime,
we will limit ourselves to the low density regime.
The dynamics of both kind of excitons has been for example discussed in
quantum wells~\cite{Thranhardt2000} and, more recently, for transition metal dichalcogenides~\cite{Selig2017,Berghauser2018}.
In these works the exciton is described using a model Hamiltonian, i.e. the Wannier equation, and its dynamics is described via
the introduction of some effective exciton--exciton or exciton--phonon interaction.
In the present work, instead, we put forward an approach which is well suited for an ab-initio (ai) implementation,
to describe the formation of coherent and non coherent excitons on ultra-fast time scale and
in the low density regime in realistic materials.

The modeling of material properties is done by describing the electronic properties
of the system, where the electronic hamiltonian includes
the many body interaction between electrons.
In first-principles approaches, the electronic problem is recast
in terms of an effective hamiltonian whose solution can be reached self-consistently.
One of the most successful example is density functional theory (DFT) where the many-body interaction is replaced by an 
effective potential describing exchange and correlation (xc) effects. The great success of DFT is due to the relative
low computational cost, within the local density approximation for the effective potential, and, yet, very high
accuracy in describing the equilibrium properties of many materials.
The approach however has some well known limitations, such as the underestimation of the electronic band gap.
Most importantly its extension to the liner response domain, i.e. time dependent (TD)-DFT~\cite{Runge1984,Ullrich-book} cannot easily capture
the physics of the exciton~\cite{Turkowski2009,Reining2002} within the common adiabatic approximations.
An approach which overcomes this limitation, at the price of higher computational cost, is MBPT.
The Bethe-Salpeter equation (BSE) of MBPT is the state-of-the-art equation for the definition of excitonic effects~\cite{Albrecth1998}. The BSE, when used on
top of Kohn--Sham states, is a fundamental brick of the Ab--Initio MBPT\,(ai--MBPT)~\cite{Onida2002}.
ai-MBPT has been indeed successfully applied to the description of the equilibrium and optical properties of
a wide range of materials, from 3D bulk semiconductors to 2D layered systems such as graphene and transition metal
dichalcogenides, 1D carbon nanotubes and complex molecules.

When dealing with pump and probe experiments in extended systems it is then natural to start from the NEQ
extension of ai-MBPT\,(ai--NEQ--MBPT) for the two following reasons. (i) The need of a correct coupling with the laser pulses, i.e.
the need to describe the physics of the exciton. (ii) The need for reliable approximations to capture NEQ xc-effects;
while MBPT offers a systematic way to introduce higher order approximations, for DFT good approximations are known mostly
for equilibrium properties. Last but not least the exploration of materials in the NEQ regime, atto-second to pico-second
time scale, is still in its infancy compared to equilibrium. It is then reasonable to focus on a more sophisticated
approach accepting the higher computational cost.
ai-NEQ-MBPT has been indeed recently implemented~\cite{Attaccalite2011,Marini2013,Perfetto2015a} and applied to both extended systems~\cite{Sangalli2015b,Sangalli2016},
2D materials~\cite{Pogna2016}, atoms and molecules~\cite{Perfetto2015b,Perfetto2018} by the authors of the present manuscript.
The Kadanoff-Baym equation (KBE) is the key equation of the approach and
describes the time evolution of the electrons in the material under the action of an external laser
pulse~\cite{kb-book,svl-book,Danielewicz1984,Stan2009,bb-book}.
Being an exact equation, it describes the coherent dynamics following the pump pulse. However it also
reduces to the semi-classical Boltzmann equation under specific approximations. It thus includes also the de-coherence process and the transition between the two regimes.
Indeed it has been shown that the KBE captures coherent excitons and that it describes the non-coherent dynamics of
electron and holes. Here we will show how the KBE can be extended to describe both regimes in the excitonic picture.

The layout of the work will be the following.
In sec.~\ref{sec:coher} we introduce the concept of fully coherent, partially coherent and non coherent states for bosons (sec.~\ref{subsec:coher-states})
and how these concepts can be linked to the polarization and population of $eh$-pairs of fermions  (sec.~\ref{subsec:coher-fermions}).
In sec.~\ref{sec:NEQ} we than discuss coherent and non coherent dynamics in the electronic system at the independent particles (IP) level (sec.~\ref{subsec:KBE})
with few results on a two band model of IP (sec.~\ref{subsec:IP_application}).
In sec.~\ref{sec:excitons} we move beyond the IP level, introducing the concept of exciton and then
considering the equation of motion (EOM) for the formation of coherent excitons (sec.~\ref{subsec:coh-exc}) and then the EOM for the formation of non-coherent excitons (sec.~\ref{subsec:coh-exc}).
We then discuss some results on a simple model (sec.~\ref{subsec:model}) and how the approach should be generalized to include 
decoherence and scattering processes in the excitonic picture (sec.~\ref{subsec:coll}).
Finally in sec.~\ref{sec:detection} we discuss how the produced (and eventually thermalized) non coherent population of excitons can
be detected in time-resolved photo-emission experiments.

%====================================================
\section{Coherent and non coherent physics} \label{sec:coher}
%====================================================

\subsection{Non coherent, partially coherent and fully coherent bosonic states}\label{subsec:coher-states}
%====================================================
We first introduce the concept of coherent~\cite{Glauber1963a,Glauber1963b} and non coherent states for bosons.
A non coherent Fock state $|n\rangle$ is defined as an eigenstate of the
particle number operator $\hat{N}=\sum_{\g} \bosdop\bosop$, while a fully coherent state $|\alpha\rangle$ is an eigenstate
of the annihilation operator $\bosop$:
\begin{eqnarray}
\label{eq:fock}
\hat{N}|n\rangle&=&n|n\rangle, \\
\label{eq:fully_coherent}
\bosop |\alpha\rangle&=&\alpha|\alpha\rangle. 
\end{eqnarray}
While Fock states are orthogonal and represents an exact basis set, coherent states are an 
over-complete basis set and are not fully orthogonal. They are also states with minimal indetermination.
The two are linked by the expression
\begin{equation}
\label{eq:coherent-fock}
|\alpha\rangle=e^{\frac{-|\alpha|^2}{2}}\sum_{n=0}^{+\infty}\frac{\alpha^n}{n!}|n\rangle.
\end{equation}
The expectation value of a fully coherent state over the number operator is finite $N=\langle\hat{N}\rangle$,
with $\alpha=\sqrt{N}e^{i\phi}$ and $\phi$ an arbitrary phase. On the contrary the expectation
value of the annihilation (or creation) operator over a Fock state is zero.
A special role is then played by the operators which are defined  as linear combinations of $\bosop$ and $\bosdop$.
Among these operators there are for example the electric field $E$ if $\bosop$ represents a photon and
the displacement of an atom from its equilibrium position $\Delta R$ if $\bosop$ represents a phonon.
In general we will refer to a state as coherent if its expectation value on such operator is non zero.
\begin{equation}
\label{eq:partially_coherent}
\langle \Psi| \bosop |\Psi\rangle\neq 0
\end{equation}
This defines a state which is at least \emph{partially coherent} and not necessarily \emph{fully coherent}
in the sense of Eq.~(\ref{eq:coherent-fock}). The quantification of coherence is an interesting topics by itself~\cite{Baumgratz2014};
for the present manuscript however the definition introduced with Eq~(\ref{eq:partially_coherent} will be sufficient.
Of course also a fully coherent state satisfy Eq.~(\ref{eq:partially_coherent}),
while a Fock state does not. A coherent state must involve, at least, the linear combination of two
Fock states with different particles number. 

\subsection{Fermion pairs and excitons: population and polarization}\label{subsec:coher-fermions}
%====================================================================
In order to bridge these concepts with the idea of coherent and non coherent electrons dynamics
we now consider the case of ``composite bosons'', i.e. linear combinations of $eh$ pairs.
Two examples are the magnon with the related coherent magnetization $M$ and the exciton with the related
coherent polarization $P$. In the present manuscript we will focus on the exciton $\bosop\ra\excop$.
To this end we introduce the electronic creation and annihilation operators in valence and conduction band
$\evdop$,$\evop$ and $\ecdop$,$\ecop$.
The population operator and the coherent polarization operator are then defined as
\begin{eqnarray}
\hat{n}^{el}&=&\sum_{cv\blk} [\evdop\evop + \ecpqdop\ecpqop], \\
\hat{P}^L(\blq)&=&\sum_{cv\blk} [d^L_{c\blk+\blq,v\blk}\ecpqdop\evop+d^L_{v\blk,c\blk+\blq}\evdop\ecpqop ],
\end{eqnarray}
with $d^L_{c\blk+\blq,v\blk}=\langle c\blk+\blq | e^{i\blq\cdot\blx} | v\blk \rangle$.
The polarization in the dipole approximation, i.e. $\blq\ra 0$, reduces to the full polarization vector
\begin{equation}
\hat{\blP}=\sum_{cv\blk} [\mathbf{d}_{cv\blk}\ecdop\evop+\mathbf{d}_{vc\blk}\evdop\ecop ] ,
\end{equation}
where the three components are obtained from the limit along three different directions.
Notice that the use of $\blq$ in the definition of the number operator has no effect, since the sum is
over all $\blk$ in the BZ. However its explicit presence makes the equation more symmetric with the one
of the polarization.
The excitonic operator can then be defined as
\begin{equation}
\excop=\sum_{cv\blk}A^{\g q}_{cv\blk} {\hat{c}_{\blk+\blq}}\evop.
\end{equation}
$A^{\l q}_{cv\blk}$ is the excitonic wave-function. We will later specify how to determine it.
In the excitonic picture the operators for populations and polarization read
\begin{eqnarray}
\hat{N}^{exc}(\blq)&=&\sum_{\g} \excdop\excop, \\
\hat{P}^L(\blq)&=&\sum_{\g} [d^{L,*}_{\g \blq}\excdop+d^{L}_{\g \blq} \excop ] .
\end{eqnarray}
We underline that in the present manuscript we have in mind the ``low density regime'' where
more than one bound exciton can be created in ``almost the same'' quantum state, as discussed in
app.~\ref{App:bound_excitons}.
Moreover the polarization operator introduced here neglects the intra-band or diagonal terms $\ecpqdop\ecop$.
This is why we refer to it as the coherent part of the polarization or the ``coherent polarization''.
The identification of the off-diagonal terms of this operator with the coherent contribution
is well defined precisely in such low density regime. We will come back to this point later.

%====================================================
\section{Non--Equilibrium dynamics} \label{sec:NEQ}
%====================================================

\subsection{Kadanoff-Baym equation for independent particles}\label{subsec:KBE}
%====================================================
We now introduce the EOM for the dynamics of an electronic system under an external laser pulse.
Within MBPT such equation is the KBE.
Within the Generalized Kadanoff Baym ansatz  (GKBA) it represents a closed equation for the one body
density--matrix of the system
\begin{equation}
\partial_t \rhom(t) - i [\hm^{eq}+\D\Sm^{s,t},\rhom(t)]-i[\hm^{ext}(q_0,t),\rhom(t)]=-\Imm(t).
\label{eq:kbe}
\end{equation}
Here $\rhom(t)=\rho_{n\blk,m\blp}(t)=\r_{l\blq}$ is the density matrix written in the basis of the eigen-functions of the equilibrium hamiltonian
$h^{eq}_{n\blk,m\blp}=\d_{n,m}\d_{\blk,\blp}\e_{n\blk}$. $\hm^{eq}$ is defined within MBPT using a quasi-particle\,(QP) approximation.
$l\blq$ is a super-index defined as ${l\blq=n\blk m\blp}$ with $\blq=\blk-\blp$.
We underline all quantities that are vectors in the $l\blq$ space (matrices in the $n\blk$ space).
\begin{equation}
[\hm,\rhom]_{n\blk,m\blp}= h_{n\blk,\nbar \kbar}\r_{\nbar \kbar,m\blp} - \r_{n\blk,\nbar \kbar}h_{\nbar \kbar,m\blp}
\end{equation}
defines the commutator (a sum for the indices with an overline ($\sum_{\nbar\kbar}$) is implicit, here as in the rest of the manuscript).
$\Delta\Sigma^{s,t}_{lq}=\Delta\Sigma^{s}_{n\blk,m\blp}[\r(t)]$ is the variation of the static ($s$) part of the self-energy,
which however depends on time ($t$) via its instantaneous functional dependence on the density--matrix.
$I_{n\blk,m\blp}(t)$ is the collision integral
which accounts for the dynamical terms of the self-energy. The GKBA enters in its construction which 
also need the expression for the retarded and advanced propagators $\Gm^{(r/a)}$.
Finally the term 
\begin{equation}
h^{ext}_{l\blq}(\blq_0,t)=\d_{\blq,\blq_0}E(t)d^L_{l\blq}
\end{equation}
is the projection of $h^{ext}(\blx,t)=E(t)e^{i\blq_0\cdot\blx}$ in the basis set of the eigenstates of $\hm^{eq}$,
and describes the interaction with an external
longitudinal electric field of modulus $E(t)$.
Here $h^{ext}_{l\blq}(\blq_0,t)=0$ if $n= m$, i.e. we neglect
intra-band transitions in the action of the external field.
Within the dipole approximation, i.e. $\blq_0\ra 0$, $\hm^{ext}$ describes an external laser pulse. 
At this level the KBE describes both coherent and non-coherent dynamics.

We can now bridge a relation between the concept of coherence introduced in sec.~\ref{subsec:coher-states}
and the density matrix by means of Eq.~(\ref{eq:partially_coherent}).
Let as first consider the case of non interacting particles, which means $\D\Sm^s=\Imm=0$.
In the present manuscript we always consider cases where the starting point ($t=0$) is a well
defined eigenstate of the system (even a single slater determinant). This is usually the assumption in
first-principles simulations on systems with a gap, i.e. that the ground state is well represented by weakly
interacting QPs at zero electronic temperature.
The expectation value of the population and the coherent polarization operators are
\begin{eqnarray}
n_{el}(t)&=&\sum_k [\rho_{v\blk,v\blk}+\rho_{c\blk+\blq,c\blk+\blq}], \\
P_L(q,t)&=&\sum_k \Big[ d^L_{c\blk+\blq,v\blk} \rho_{c\blk+\blq,v\blk}(t) + d^L_{v\blk,c\blk+\blq} \rho_{v\blk,c\blk+\blq}(t)\Big],
\end{eqnarray}
i.e. the diagonal elements of $\r$ define the populations $f_{n\blk}=\rho_{n\blk,n\blk}$,
while the off-diagonal matrix elements define the polarization.
{\em A state is at least partially coherent if some $\rho_{n\blk,m\blp}\neq 0$ for $n\neq m$, while the
density--matrix becomes diagonal once decoherence is completed}.

We can also disentangle the coherent and the non-coherent dynamics expanding the KBE
to second order in the external field. In this way we adopt  the ``low density'' regime by assuming the external field is weak, i.e. we are within the 
``low pumping'' regime:
\begin{eqnarray}
\label{eq:KBE_coh_IP}
\partial_t \rhom^{(1)}(t) - i [\hm^{eq},\rhom^{(1)}]-i[\hm^{ext}(t),\rhom^{eq}]=0, \\
\label{eq:KBE_pop_IP}
\partial_t \rhom^{(2)}(t) - i [\hm^{eq},\rhom^{(2)}]-i[\hm^{ext}(t),\rhom^{(1)}]=0.
\end{eqnarray}
To linear order in the external field, only the off diagonal terms change with time
(this result can be proved also in case static correlations are considered, i.e. $\D\Sm^s\neq 0$),
while $\r^{(1)}_{nn\blk}=0$. Thus Eq.~\eqref{eq:KBE_coh_IP} is the EOM for the coherences.
Eq.~\eqref{eq:KBE_pop_IP} is the EOM for the populations if the terms $\r^{(2)}_{n\blk,n\blk}$ only are considered.

We have thus obtained that the separation in coherences and populations can
be achieved by expanding the KBE in the IP case. Notice that such separation has a straightforward 
interpretation: the external pulse first creates coherent eh-pairs, some of which, further interacting with the external
field, become non-coherent eh-populations.

\subsection{Results for an infinite system of non interacting particles}\label{subsec:IP_application}
%====================================================
To get further insight we assume the equilibrium hamiltonian describes an infinite
system with one fully occupied valence band and one empty conduction band.
For non-interacting particles $\rhom$ defines all physical properties of the system
and in particular the property $\rhom=\rhom^2$ defines if the system is in a pure state or not, i.e. if there
exist a wave function, coherent or not, which is related to the density matrix.
In this case we can even directly write the wave-function
associated with a given density matrix.

The 2x2 equilibrium density--matrix for each $\blk,\blp$ pair is
\begin{equation}
\label{eq:rho_ip_eq}
\rhom_{\blk\blp}:=
\begin{pmatrix}
\rho_{v\blk,v\blk} & \rho_{v\blk,c\blp} \\
\rho_{c\blp,v\blk} & \rho_{c\blp,c\blp}
\end{pmatrix}
=
\delta_{\blk,\blp}
\begin{pmatrix}
1 & 0 \\
0 & 0
\end{pmatrix},
\end{equation}
which corresponds to the ground state $|\psi_g\rangle=\prod_\blk \evdop |0\rangle$.
The external pump pulse will make the system evolve in a coherent way,
sending it in a coherent superposition of states with electrons excited from $\blk$ to $\blk+\blq_0$.
To linear order in the external field, i.e. assuming an expansion of the wave-function in terms
of single excitations, we can write
\begin{equation}
\label{eq:part_coher_ip}
|\Psi(t)\rangle=\sqrt{\frac{\Omega_{BZ}}{n_\blk}}\sum_\blk\big[\sqrt{1-\fkq(t)}+\sqrt{\fkq(t)}e^{i\D\ekq t}\ecpqdop\evop\big]e^{iE_gt}|\Q_g\rangle.
\end{equation}
with $\fkq(t)\propto |E|^2$, $\blq=\blq_0$ and $\D\ekq=\e_{c\blk+\blq}-\e_{v\blk}$.
$n_\blk$ is the number of $\blk$--points and $\Omega_{BZ}$ the size of the Brillouine zone (BZ).
Here we introduced $\fkq(t)$ as a coefficient of the wave-function. We will immediately show,
building the associated density matrix, that it indeed defines the electronic occupations.

Beside the time evolution, there is a main difference between
the fully coherent state introduced in Eq.\eqref{eq:coherent-fock} and Eq.\eqref{eq:part_coher_ip}:
the sum in the latter is truncated to one $eh$ pair. This is due to our assumption for the structure of the
wave-function (see also App.~\ref{App:coherent_states}).
However we also need to remark that any term involving two $eh$-pairs should involve different $\blk$-points since, due
to Pauli exclusion principle, multiple $eh$ pairs in the same state are not allowed.
As a consequence it is not possible to construct the analogous of Eq.\eqref{eq:coherent-fock}.
Indeed a single eh-pair is far from being a ``quasi-boson''. This will not be a limit in the correlated case,
where, at low pumping, an excitons is a good ``quasi-boson'', i.e. it is possible to fill with more than one excitons
almost identical states (see also appendix~\ref{App:bound_excitons}).

The density--matrix corresponding to Eq.~\eqref{eq:part_coher_ip} is different from zero only for blocks with $\blk=\blp+\blq_0$:
\begin{equation}
\label{eq:rho_part_coh_ip}
\rhom_{\blk\blp}=\d_{\blk+\blq_0,\blp}
\begin{pmatrix}
1-\fkq(t) & \sqrt{(1-\fkq(t))\fkq(t)}e^{i\D\ekq t} \\
\sqrt{(1-\fkq(t))\fkq(t)}e^{-i\D\ekq t} & \fkq(t)
\end{pmatrix}.
\end{equation}
Here $\rhom^2=\rhom$ at each time.
It is possible to show that the first order expansion of the off-diagonal elements 
$\rho^{(1)}_{c\blk+\blq_0,v\blk}(t)=\sqrt{\fkq(t)}e^{i\D\ekq t}$ is solution of Eq.~\eqref{eq:KBE_coh_IP}
while the diagonal terms, $n=m$, are solution of Eq.~\eqref{eq:KBE_pop_IP}.
Since $I(t)=0$, the occupations $\fkq(t)$ do not evolve anymore for $t>t_f$,
i.e. when $\hm^{ext}=0$. If we then allow the the system to loose coherence the density--matrix becomes
\begin{equation}
\rhom_{\blk\blp}(t)=\d_{\blk+\blq_0,\blp}
\begin{pmatrix}
1-\fkq(t_f) & 0 \\
0 & \fkq(t_f)
\end{pmatrix}.
\end{equation}
Notice that this final state cannot be represented anymore as a wave-function since $\rho$ is not pure,
i.e. $\rhom\neq\rhom^2$.
It is a non-coherent superposition of Fock states with populations $f_{v\blk}(t_f)=1-\fkq(t_f)$ and $f_{c\blk+\blq_0}(t_f)=\fkq(t_f)$.
Accordingly $\fkq$ defines the electrons removed from $v\blk$ and promoted to $c\blk+\blq$.
The assumption of decoherence, here introduced ad-hoc, turns the description in terms of the density--matrix from
deterministic (in the sense that the density--matrix can be associated to an existing wave-function which
evolves in a deterministic way, i.e. describes what is called a pure state) to statistical.

%====================================================
\section{Excitons Out--of--equilibrium} \label{sec:excitons}
%====================================================

\subsection{Coherent excitons}\label{subsec:coh-exc}
%====================================================
We now want to turn our attention to the description of excitons.
To this end we need to activate the change in the static part of the self-energy $\D\Sm^s=\D\Sm^{HSEX}$
which contains the variation of the Hartree plus Screened Exchange (HSEX) self-energy.
At this point we have to observe that, allowing for a change in the self-energy, we allow the hamiltonian
to evolve in time. Accordingly the basis-set which instantaneously diagonalizes the Hamiltonian will evolve in time as well.
The diagonal and off-diagonal matrix elements on such instantaneous basis-set will both be a mixture
of the diagonal and off-diagonal matrix elements in the equilibrium basis set. This would introduce a significant complication
in the attempt to distinguish coherent and non coherent terms in the density--matrix. Moreover it would bring us to have the
excitonic operator which evolves in time (since the excitonic hamiltonian would be time dependent as well).
To avoid all these complications we consider the case where a finite number of excitons is created.
A finite number of excitons in an infinite system means the changes in the density--matrix
are infinitesimal. This situation also describes,
to a good degree of approximation, the low pumping regime where the exciton densities is low.
Then the QP basis set (and the excitonic operator we are going to introduce) can be kept static.
We can thus keep the distinction between coherences (off-diagonal terms of the density--matrix) and populations
(diagonal terms of the density--matrix). 

If the KBE is linearized in the external field and the HSEX self-energy is used
together with the QP-GW approximation for the equilibrium hamiltonian,
it reduces to the  Bethe-Salpeter equation (BSE) which is known to well describe excitons in
extended systems at equilibrium~\cite{Attaccalite2011}. We will refer to this as the TD-HSEX approach.
Since to linear order in the external field, only the off diagonal terms
of the density--matrix are changed by the external field, the TD-HSEX describes indeed coherent excitons.
The linearized TD-HSEX equation for the off diagonal elements of $\rho$ reads
\begin{equation}
\partial_t \rhom^{(1)}(t) - i [\hm^{eq},\rhom^{(1)}]-i[\Kmm^{HSEX} \d\rhom^{(1)},\rhom^{eq}]-i[\hm^{ext}(t),\rhom^{eq}]=0,
\end{equation}
where we have introduced $K^{HSEX}_{l\blq,l'\blq'}=\d\S^{HSEX}_{l\blq}/\d\r_{l'\blq'}\big|_{\rhom^{eq}}$.
$\Kmm$ is a matrix in the super-indexes space.
Notice that, although $\rhom^{(1)}$ is infinitesimal, $\Kmm^{HSEX}$ is not.
Using the expression for $\hm^{eq}$ and $\rhom^{eq}$ of Eq.~\eqref{eq:rho_ip_eq} we can introduce the excitonic Hamiltonian
\begin{equation}
H^{exc}_{lq,l'q'}(\blq)=(\epsilon_{c\blk+\blq}-\epsilon_{v\blk})\delta_{c,c'}\delta_{v,v'}\delta(\blq-\blq')+K^{HSEX}_{l\blq,l'\blq'}, \\
\end{equation}
and diagonalize it defining the excitonic eigen-states (or wave-functions)
$A^{\g\blq}_{cv\blk}$ and eigen-energies  $E_\g(\blq)$.
The operator which creates an exciton is 
\begin{equation}
\excdop=\sum_\blk A^{\g\blq}_{cv\blk} \ecpqdop\evop.
\label{eq:eq_bse}
\end{equation}
If we now rotate the Eq.\eqref{eq:kbe} in the excitonic basis-set defined by Eq.\eqref{eq:eq_bse}:
\begin{equation}
\label{eq:eom-coh-exc}
\partial_t \rho^{(1)}_{\g\blq}(t) - i E_{\g\blq}\rho^{(1)}_{\g\blq}(t)=ih^{ext}_{\g\blq}(\blq_0,t).
\end{equation}
$\r_{\g\blq}$ defines coherent excitons explicitly and Eq.~\eqref{eq:eom-coh-exc} describes their creation.
We immediately notice however that, at variance with the IP case,
the one-body density--matrix only describes
coherent excitons, i.e. it cannot be used to describe non-coherent exciton populations.
In the excitonic picture polarization (or coherences) on one hand and 
populations (or non coherent Fock states) on the other hand are associated with two operators
which are different in nature. To describe populations,
the expectation value of $\excdop\excop$, i.e. the two--body density--matrix, is best suited.
This has been of course already remarked in the literature~\cite{Koch2006}.
We here underline that the KBE is an exact equation. Accordingly the physics of non-coherent excitons 
can be  in principle captured also within the one--body density--matrix.
However even the exact one--body density--matrix contains less informations compare to the two--body
one and the number of excitons in the system cannot be extracted.
In Ref.~\cite{Perfetto2016} we discussed how the one-body density--matrix can be used to describe the
signature of non-coherent excitons in photo-emission, provided the correct choice for the self-energy is done.
Staying within the one body density--matrix however is highly non trivial and
calls for the need of the correlated T-matrix approximation to the self-energy. This is similar, in some sense, to
the difficulties one encounters in defining excitons at equilibrium within the two point
response function of the Heidin equation, where non trivial vertex corrections need to be included.
On the contrary introducing the four point response function $L$, excitons are well described considering the
static $HSEX$ kernel.

\subsection{Non coherent excitons}\label{subsec:non-coh-exc}
%====================================================
We thus turn our attention to the two--body density--matrix. To write its equation of motion we start from
the electron hole propagator on the contour $\Lmm$ and write its Dyson equation in presence of the static HSEX kernel
\begin{equation}
\Lmm(t,t')=\Ltmm^0(t,t')+\Ltmm^0(t,\tbar) \Kmm^{HSEX} \Lmm (\tbar,t'),
\end{equation}
with $\Ltmm^0=\Gtm\times\Gtm$ with the indexes as follow
\begin{equation}
\Lt^0_{n\blk n'\blk',m\blp m\blp'}(t,t')=\Gt_{n\blk,m\blp}(t,t')\Gt_{n'\blk',m'\blp'}(t',t)
\end{equation}
and $\Gt$ the one body Green function in presence of an external potential:
\begin{equation}
\Gt^{-1}_{n\blk,m\blp}(t,t')=G^{-1}_{n\blk,m\blp}(t,t')-h^{ext}_{n\blk,m\blp}(\blq_0,t)\d(t,t').
\end{equation}
Defining $\Lmm^0=\Gm\times\Gm$, the Dyson equation becomes
\begin{equation}
\label{eq:DysonL}
\Lmm(t,t')=\Lmm^0(t,t')+\Lmm^0(t,\tbar)
  \big[\Kmm^{HSEX}\d(\tbar-\tbar')+\Kmm^{ext}(\tbar,\tbar') \big] \Lmm(\tbar',t'),
\end{equation}
with
\begin{multline}
K^{ext}_{n\blk n'\blk',m\blp m'\blp'}(t,t')=
G^{-1}_{n\blk,m\blp}(t,t')h^{ext}_{n'\blk',m'\blp'}(\blq_0,t)\d(t-t') \\
 +h^{ext}_{n\blk,m\blp}(\blq_0,t)\d(t-t')G^{-1}_{n'\blk',m'\blp'}(t',t) \\
  +h^{ext}_{n\blk,m\blp}(\blq_0,t)h^{ext}_{n'\blk',m'\blp'}(\blq_0,t)\d(t-t')
\end{multline}
describing the three processes where the external perturbation 
acts on the conduction electron only $\Kmm^{ext,1c}$, on the valence electron only $\Kmm^{ext,1v}$ or, finally, on both $\Kmm^{ext,2}$.
We want to focus on the EOM for the terms $L_{cv,cv}:=L_{c\blk+\blq v\blk+\blq,c'\blk'+\blq'v'\blk'}(\w)$
in a two band model.
We notice, however, that the equation cannot be closed for such terms only
since $\Kmm^{ext}$ sends $c\ra v$ and $v\ra c$ due to the off-diagonal structure in $cv$ space of $\hm^{ext}$.
Instead, since we are interested in the description of excitons, we
consider $K^{HSEX}$ different from zero, and thus $L\neq G\times G$, only in the $cv,cv$ channel.
The terms which are sent to $cv,cv$ from the action of $K^{ext,1i}$ are $L_{cc,cv}$, $L_{vv,cv}$, $L_{cv,vv}$
and $L_{cv,cc}$. If $\r_{n\blk,m\blp}^{eq}=\d_{\blk,\blp}\d_{n,v}\d_{n,m}$
and having assumed $K^{HSEX}=0$ outside the $cv,cv$ channel, even the correlated $L$ for such terms
will be $L= G\times G$.
For example
\begin{equation}
\label{eq:Gamma_rho_rel_general}
L_{c\blk c\blk',c\blp v\blp'}(t,t')\,=\, G_{c\blk c\blp}(t,t') G_{c\blk'v\blp'}(t',t).
\end{equation}
To define the EOM for $L^{{}<{}}$ we rewrite Eq.~(\ref{eq:DysonL}) in the form $(L_0^{-1}-K)L=1$
and $L(L_0^{-1}-K)=1$. The terms involving $K^{ext}$ thus become:
\begin{eqnarray}
K^{ext,1v}_{c\blk v\blk',c\kbar c\kbar'}(t,\tbar) L_{c\kbar c\kbar',c \blp v \blp'}(\tbar,t')=
  h^{ext}_{v\blk',c\kbar'}(\blq_0,t) G_{c\kbar',v\blp'}(t,t')\d_{\blk,\blp}, \\ %\d(t-\tbar) \\
K^{ext,1c}_{c\blk v\blk',v\kbar v\kbar'}(t,\tbar) L_{v\kbar v\kbar',c \blp v \blp'}(\tbar,t')=
  h^{ext}_{c\blk,v\kbar}(\blq_0,t) G_{v\kbar,c\blp}(t,t')\d_{\blk',\blp'}, \\ %\d(t-\tbar) \\
L_{c\blk v\blk',c\kbar c\kbar'}(t,\tbar) K^{ext,1v}_{c\kbar c\kbar',c \blp v \blp'}(\tbar,t')=
  G_{v\blk',c\kbar'}(t,t') h^{ext}_{c\kbar',v\blp'}(\blq_0,t')\d_{\blk,\blp}, \\ %\d(t-\tbar) \\
L_{c\blk v\blk',v\kbar v\kbar'}(t,\tbar) K^{ext,1c}_{v\kbar v\kbar',c \blp v \blp'}(\tbar,t')=
  G_{c\blk,v\kbar}(t,t') h^{ext}_{v\kbar,c\blp}(\blq_0,t')\d_{\blk',\blp'}. %\d(t-\tbar) \\
\end{eqnarray}
Let us label the sum of these four terms as $H^{ext}_{c\blk v\blk',c\blp v\blp'}[G(t,t')]$.
For the $\Kmm^{ext,2}$ we need to consider the terms $L_{vc,cv}$, $L_{cv,vc}$.
However since we want to consider terms up to second order in the external field
we need terms of the kind $\Kmm^{ext,2} \Lmm^{(eq)}$ which are identically zero for the indices just
considered.
Having replaced the terms involving $\Kmm^{ext}\Lmm$ with a 
term $\Hmm^{ext}$ which is independent from $\Lmm$ in Eq.~\eqref{eq:DysonL}, we can
easily move from the contour to the real time axis, thus introducing
the EOM for the electron--hole two--body density--matrix, $\Gmm^{eh}$, up to second order in the field.
We define $\Gmm^{eh}$ as
\begin{equation}
\G^{eh}_{c\blk v\blk',c\blp v\blp'}(t)=L^{{}<{},eh}_{c\blk v\blk',c\blp v\blp'}(t,t)=\langle \hat{c}^\dag_{\blk}(t)\hat{v}_{\blk'}(t)\hat{c}_{\blp}(t)\hat{v}^\dag_{\blp'}(t) \rangle.
\end{equation}
We will omit the $eh$ suffix from now on.
By using $(L^0)^{-1}_{cv\blk,c'v'\blk'}(q)=i\partial_t-(\epsilon_{c\blk+\blq}-\epsilon_{v\blk})$, the fact that
the $\Kmm^{HSEX}$ is static, and the resulting structure of the $\Kmm^{ext}$ term, we obtain
\begin{equation}
\partial_t \Gmm^{(2)}(t) -i[\Hmm^{exc},\Gmm^{(2)}]  - i [\hm^{ext},\rhom^{(1)}]\times\dm- i\dm\times [\hm^{ext},\rhom^{(1)}] =0.
\end{equation}
We now rotate in the equilibrium excitonic basis and
consider only the diagonal terms which define the excitonic populations $N_{\g\blq}=\G_{\g\blq,\g\blq}$.
The term $[\Hmm^{exc},\Gmm]=0$ for $\g=\g'$ since $\Hmm^{exc}$ is diagonal in excitonic space.
Taking into account the EOM for the one body density--matrix, we finally have
\begin{eqnarray}
\label{eq:KBE_coh_EXC}
\partial_t \rho^{(1)}_{\g\blq}(t) -i E_{\g\blq}\rho^{(1)}_{\g\blq}(t)-i h^{ext}_{\g\blq}(\blq_0,t)&=& 0, \\
\label{eq:KBE_pop_EXC}
\partial_t \G^{(2)}_{\g\blq,\g\blq}(t) -i H^{ext}_{\g\blq\g\blq}[\r(t)] &=& 0.
\end{eqnarray}
The coupled Eqs.~\ref{eq:KBE_coh_EXC}-\ref{eq:KBE_pop_EXC} constitute a first result of the present 
manuscript and could be easily implemented from first-principles. They are the correlated version of
Eqs.~\eqref{eq:KBE_coh_IP}-\eqref{eq:KBE_pop_IP}.
Only the knowledge of the excitonic Hamiltonian is required.
Once an external perturbation $\hm^{ext}(\blq_0,t)$ is selected, $\Hmm^{exc}$ needs to be diagonalized only for $\blq=\blq_0$.
A laser pulse is described by using the optical limit $\blq_0\ra 0$.
The EOM, within the static HSEX approximation, does not mix terms with different $\blq$.
It is worth to observe that an exciton population $N_{\g\blq}$ is not directly related
to a coherent exciton $\r_\g$, i.e. we do not obtain a term of the form $h^{ext}_{\g\blq}\,\r_{\g\blq}$,
but rather terms of the form
\begin{eqnarray}
H^{ext}_{\g\blq\g\blq}[\r(t)]&=&\sum_{\blk,\blk'} A^{\g\blq,*}_{cv\blk} H^{ext}_{c\blk+\blq v\blk,c\blk'+\blq v\blk'}[\r(t)] A^{\g\blq}_{cv\blk'},
\end{eqnarray}
where a summation of coherent $eh$ pairs appears. 
This is a manifestation of the composite
nature of the exciton.
We underline that Eq.~\eqref{eq:KBE_pop_EXC} requires the knowledge of $\r_{n\blk,m\blp}$ and not only
of $\r_{\g\blq}$. However the former can be easily obtained from the latter by a rotation back from the
correlated to the IP basis set.

\subsection{Two--bands model in the independent--particles case}\label{subsec:model_IP}
%====================================================
Let us investigate what happens in the IP level
for the two bands case considered before and using the linearized expression for
the wave-function of Eq.~\eqref{eq:part_coher_ip}.
At equilibrium we easily get $\G^{eq}_{c\blk v\blk',c\blp v\blp'}=0$, 
i.e. the excitons population is zero.
Once the laser pulse is switched on we obtain
\begin{equation}
\G_{c\blk v\blk',c\blp v\blp'}(t)=\delta_{\blk+\blq_0,\blk'}\delta_{\blp+\blq_0,\blp'}\sqrt{\fkq(t)\fkpq(t)}e^{i(\D\ekq-\D\ekpq)t}.
\end{equation}
The coherences described by the off-diagonal elements of $\Gmm$ involve two $eh$
pairs at $\blk$ and $\blk'$ and are not directly related to the coherences described by $\rho$.
Indeed for all terms different from zero $\Gmm(t)\propto E^2$.
We now consider the four possible terms entering $[\Hmm^{ext},\Gmm]$.
They all vanish at equilibrium, whereas out of equilibrium only two are different from zero:
\begin{eqnarray}
\G_{c\blk v\blk',c\blp c\blp'}=\delta_{\blk+\blq_0,\blk'}\delta_{\blp,\blp'}\sqrt{(1-\fkpq(t))\fkq(t)}e^{-i(\D\ekpq)}, \\
\G_{v\blk v\blk',c\blp v\blp'}=\delta_{\blk,\blk'}\delta_{\blp+\blq_0,\blp'}\sqrt{(1-\fkq(t))\fkpq(t)}e^{i(\D\ekq)}.
\end{eqnarray}
These terms are directly related to the coherent excitons.
The first order expansion can be expressed in terms of $\rho$:
\begin{eqnarray}
\label{eq:Gamma_rho_rel1_IP}
\G^{(1)}_{c\blk v\blk',c\blp c\blp'}(t)=\rho^{(1)}_{c\blk,v\blk'}(t)\overline{\rho}^{eq}_{c\blp,c\blp'}, \\
\label{eq:Gamma_rho_rel2_IP}
\G^{(1)}_{v\blk v\blk',c\blp v\blp'}(t)=\rho^{eq}_{v\blk,v\blk'}\overline{\rho}^{(1)}_{c\blp,v\blp'}(t).
\end{eqnarray}

\subsection{Two--bands model for the correlated case}\label{subsec:model}
%====================================================
%
We then consider a minimal model for excitons, recently introduced to discuss excitonic signature in 
photo-emission~\cite{Perfetto2016}
\begin{eqnarray}
\hat{H}_{\rm ins}&=&\sum_{\blk}(\e_{v\blk}\evdop\evop
+\e_{c\blk}\ecdop\ecop)-
U(0)\frac{N^{el}_{v}}{\callL}\sum_{\blk}\ecdop\ecop
\nn\\
&+&
\frac{1}{\callL}\sum_{\blk_{1}\blk_{2}\blq}U(q)\,\hat{v}^{\dag}_{\blk_{1}+\blq}\hat{c}^{\dag}_{\blk_{2}-\blq}
\hat{c}_{\blk_{2}}\hat{v}_{\blk_{1}},
\label{eq:Hmodel}
\end{eqnarray}
and solve the equations discussed previously. Here $\callL$ is the length of the 1D model,
$N^{el}_v$ the total number of electrons in valence and $U(q)$ the interaction.
In this model electrons in the valence band interact only with electrons in the conduction band.
As a consequence the ground state of the system is the same as in the IP case: $|\Q_{g}\ket=\prod_{\blk}\evdop|0\ket$.
The second term in the hamiltonian has the role of neutralizing the $\blq=0$ repulsion exerted from
the valence electrons onto any electron sitting in conduction, similarly to what a uniform positive
background would do in the case of fully interacting electrons. 
Notice that such model is physically meaningful as long as the density of excited electrons
is negligible or very low, i.e. the situation on which we focus in the present manuscript.

We now consider the excited states obtained from the linear combination of single particle excitations.
As discussed in Ref.~\cite{Perfetto2016} the resulting excitonic wave-function, eigenstate of the
hamiltonian $\hat{H}_{\rm ins}$,
can be found assuming an interaction constant in momentum space $U(q)=U$
(i.e. a contact interaction in real space) as a solution of the equation
\begin{equation}
(\D\ekq-E_{\g\blq})A^{\g\blq}_{k} = \frac{U}{\callL}\sum_{\blq'\neq 0} A^{\g\blq}_{\blk+\blq'}.
\end{equation}
Here we write the general equation at finite $\blq$.
The excitonic eigenvectors take the form
\begin{equation}
\label{eq:exc_expression_model}
A^{\g\blq}_{k}=\frac{\sqrt{R_q}}{\D\ekq-E_{\g\blq}}.
\end{equation}
The related eigen-energies can be written as $E_{\g\blq}=\e_{gap,\blq}-b_{\g\blq}$. There exist a bound solution $\g=x$ with
binding energy $b_{x\blq}>0$, plus a continuum of solutions with $b_{\g\blq}<0$.
Inserting this into Eq.~\eqref{eq:exc_expression_model} and writing $\D\ekq=\e_{gap,\blq}+a_{\blk\blq}$,
with $a_{\blk\blq}\geq 0$, we obtain at the denominator $a_{\blk\blq}+b_{\g\blq}$. Since $b_{x\blq}>0$, $A^{x\blq}\propto \frac{1}{\callL}$ and the 
excitonic wave-function is fully de-localized in $k$ space as $\callL\ra\infty$ for the bound solution.
All other solutions are instead localized, since $A^{\g\blq}$ is dominated by the divergence $a_{\blk\blq}+b_{\g\blq}=0$ for $\g\neq x$.
The value of $R_{x\blq}$ and $b_{x\blq}$ are fixed by the two equations
\begin{eqnarray}
\sum_k |A^{x\blq}_{k}|^2&=&1, \\
\sum_{\blq'\neq 0} (a_{\blk\blq}+b_{x\blq'})^{-1}&=&\frac{\callL}{U}.
\end{eqnarray}
Let us first consider the ``partially coherent'' exciton state
\begin{equation}
\label{eq:part_coher_exc}
|\Psi_{1}(\blq,t)\rangle=\sum_\g (\sqrt{1-N_{\g\blq}(t)} + \sqrt{N_{\g\blq}(t)} e^{iE_{\g\blq}t}\excdop )e^{iE_gt}|\Psi_g\rangle.
\end{equation}
Here we introduce $N_{\g\blq}(t)$ as a mixing coefficient of the two states, similarly to how we introduced
$\fkq(t)$ in Eq.~(\ref{eq:part_coher_ip}). We obtain
\begin{eqnarray}
\label{eq:rho_part_coh_exc}
\rho_{\g\blq}(t)&=&   \delta_{\blq,\blq_0} \sqrt{(1-N_{\g\blq}(t))N_{\g\blq}(t)}e^{iE_{\g\blq}t}, \\
\label{eq:rho2_part_coh_exc}
\G_{\g\blq,\g\blq}(t)&=& \delta_{\blq,\blq_0} N_{\g\blq}(t).
\end{eqnarray}
Comparing the IP case, $\rho_{\g\blq}$ now plays the role that was played by off-diagonal elements $\rho_{c\blk+\blq,v\blk}$
in Eq.~\eqref{eq:rho_part_coh_ip}, while $\G_{\g\blq,\g\blq}$ the role played by the diagonal elements $\rho_{c\blk+\blq,c\blk+\blq}$.
Thus $N_{\g\blq}(t)$ defines here the excitonic populations, similarly to how $\fkq$ defined the excited electronic population.
Next we have
\begin{eqnarray}
\G_{\g\blq,c\blk c\blk'}=\delta_{\blk,\blk'}\delta_{\blq,\blq_0}\sqrt{(1-N_{\g\blq}(t))N_{\g\blq}(t)}e^{iE_{\g\blq}t}, \\
\G_{v\blk v\blk',\g'\blq'} =\delta_{\blq',\blq_0}\delta_{\blk,\blk'}\sqrt{(1-N_{\g'\blq}(t))N_{\g'\blq}(t)}e^{-iE_{\g'\blq}t},
\end{eqnarray}
which represents the correlated version of Eqs.~\eqref{eq:Gamma_rho_rel1_IP}-\eqref{eq:Gamma_rho_rel2_IP}.
For the bound excitonic peak, we can consider states of the form $(\excxdop)^n|\Q_g\rangle$
also with $n>1$ (see App.~\ref{App:bound_excitons}).
The same is not possible for non bound states with $b_{\g\blq}<0$.
The crucial difference between the two is 
the localization, in $\blk$-space, of the excitonic wave-function $A^{\g\blq}_k$.
We can thus consider the ``fully coherent'' exciton state
\begin{equation}
\label{eq:fully_coher_exc}
|\Psi_{\infty}(\blq,t)\rangle= e^{\frac{-|\alpha(t)|^2}{2}}\sum_{n=0}^{+\infty}\frac{\a^n(t)(\excxdop)^n e^{inE_{x\blq}t}}{n!} e^{iE_gt} |\Psi_g\rangle.
\end{equation}
To evaluate $\r$ and $\G$ we use the fact that $|\Psi_{\infty}(q,t)\rangle$ is an eigenstate of $\excxop$,
with eigenvalue $\alpha(t)=\sqrt{N(t)}e^{i\phi_{x\blq}}$, and
observe that at finite time a factor $e^{iE_{x\blq}t}$ must remain in $\r$:
\begin{eqnarray}
\label{eq:rho1_fully_coher_exc}
\rho_{\g}(\blq,t)&=&   \delta_{\blq,\blq_0}\delta_{\g,x} \sqrt{N_{x\blq}(t)} e^{i(E_{x\blq}t+\phi_{x\blq})}, \\
\label{eq:rho2_fully_coher_exc}
\G_{\g\blq,\g\blq}(t)&=& \delta_{\blq,\blq_0} \delta_{\g,x} N_{x\blq}(t).
\end{eqnarray}
Similarly, to evaluate the other terms of the density--matrix we simply observe that all Fock states composing the 
coherent state are eigenstates, in the low density limit (i.e. neglecting deviation which involve one single k-point) of both $\ecop\ecdop$
and $\evdop\evop$ for any $\blk$.
This gives
\begin{eqnarray}
\G_{\g\blq,c\blk c\blk'}=\delta_{\blk,\blk'}\delta_{\blq,\blq_0}\delta_{\g,x}\sqrt{N_{xq}(t)}e^{i(E_{x\blq}t+\phi_{xq})}, \\
\G_{v\blk v\blk',\g'\blq'} =\delta_{\blq',\blq_0}\delta_{\blk,\blk'}\delta_{\g,x}\sqrt{N_{xq'}(t)}e^{-i(E_{x\blq'}t-\phi_{xq})},
\end{eqnarray}
i.e., again these terms of the two--body density--matrix can be written in terms of coherent excitons.
It is again possible to show that the density matrices of Eqs.~\eqref{eq:rho1_fully_coher_exc}-\eqref{eq:rho2_fully_coher_exc}
are solutions of the excitons EOMs ,
i.e. Eqs.~\eqref{eq:KBE_coh_EXC}-\eqref{eq:KBE_pop_EXC}.
Let us underline that to construct the ``fully coherent'' state we have used the fact that the bound exciton
has some boson like properties.
The summation in the coherent state is not limited to states with a single $eh$ pair.
As a consequence, while the electronic population $f_{n\blk}(t)$ were limited by the Pauli exclusion principle,
this limitation is absent here.

\subsection{Excitonic decoherence and termalization}\label{subsec:coll}
%====================================================

Up to now we just considered the case $\Imm(t)=0$. Let us now comment on its effect on exciton dynamics, again starting 
from the non correlated case. The distinction between coherent terms $\rho^{(1)}_{n\blk,m\blp}$ with $n\neq m$ and
non coherent terms $\rho^{(2)}_{n\blk,n\blk}=f_{n\blk}$ made for the $\D\Sm^s=0$ case, leads
to a great simplification of the collision integral.
The off diagonal terms, $I_{n\blk,m\blp}(t)$ with $n\neq m$, describe the de-coherence process and can 
be approximated as $I_{n\blk,m\blp}(t)=-\eta\r_{n\blk,m\blp}(t)$. The diagonal terms describe the scattering
processes between populations and can be derived within the approximation
$I_{n\blk,n\blk}[\r](t)\approx I_{n\blk,n\blk}[f_{n'\blk'}](t)$,
used together with the GKBA, the Markovian approximation, and the assumption
that $G^{(r/a)}\approx e^{\pm i(\e_{n\blk}\pm i\g_{n\blk})t}$, i.e. a QP like structure with finite lifetime $\g_{n\blk}$.
Using all this in Eqs.~\eqref{eq:KBE_coh_IP}-\eqref{eq:KBE_pop_IP},
and writing the second equation for the diagonal terms $f_{n\blk}$ only, we obtain
\begin{gather}
\label{eq:KBE_coh_IP_scatt}
\partial_t \rho^{(1)}_{n\blk,m\blq}(t) - i \Delta\e_{n\blk,m\blp}\rho^{(1)}_{n\blk,m\blp}(t)-i h_{n\blk,n\blp}^{ext}(\blq_0,t)=-\eta\r^{(1)}_{n\blk,m\blp}(t),\\
\label{eq:KBE_pop_IP_scatt}
\partial_t \rho^{(2)}_{n\blk,n\blk}(t)-i [\hm^{ext}(t),\rhom^{(1)}]_{n\blk}=-\sum_{n'\blk'} \alpha_{n\blk,n'\blk'}\big[ (1-f_{n'\blk'}) f_{n\blk}+ f_{n'\blk'} (1-f_{n'\blk'})\big].
\end{gather}
The resulting $I(t)$ has the same structure of the semi-classical Boltzmann equation~\cite{Marini2013,Sangalli2015a}
describing scattering processes where the IP energies are conserved. 
The precise structure of $\alpha_{n\blk,n'\blk'}$
and the energy conservation factors embodied in it depend on the choice of the self-energy.
For example the second Born approximation gives that the energy $\e_{n\blk}-\e_{m\blp}$ is transferred
to another $eh$ pair $\e_{n'\blk'}-\e_{m'\blp'}$ and the process is weighted by the matrix element
of the bare electron-electron interaction. Within the $GW$ approximation
a similar process is weighted by the screened electron-electron interaction. If instead the Fan electron-phonon
self-energy is used, the energy is transferred to a phonon $\w_{\g\blq}$ and the process is weighted by the 
electron-phonon matrix elements.
Given the structure of $I(t)$, two Fermi distributions, one in valence and the other in conduction,
are stationary solutions of the EOM. Only including the
electron-photon self-energy, i.e. radiative recombination processes, the energy is
transferred to photons allowing the system to relax back to equilibrium.

In the correlated case, $\D\Sm^s=\D\Sm^{HSEX}$, similar approximations can be designed in the low pumping regime
where again coherent and non-coherent terms can be separated. We can then assume a de-coherence term of the form 
$I_\g^q(t)=-\eta\r_\g^q(t)$ and a scattering term functional of the sole excitonic populations:
\begin{gather}
\label{eq:KBE_coh_EXC_scatt}
\partial_t \rho^{(1)}_{\g\blq}(t) -i E_{\g\blq}\rho^{(1)}_{\g\blq}(t)-i h^{ext}_{\g\blq}(\blq_0,t)= -\eta\r_{\g\blq}(t), \\
\label{eq:KBE_pop_EXC_scatt}
\partial_t \G^{(2)}_{\g\blq,\g\blq}(t) -iH^{ext}_{\g\blq,\g\blq}[\r(t)] = I_{\g\blq,\g\blq}[N_{\g\blq}].
\end{gather}
Since we are describing the exciton dynamics, we want that the excitonic energies are conserved in the scattering processes.
For the case of our model, where one bound exciton $E_x(q)$ exist, well separated from the continuum,
we can derive the shape of such collision integral under a number of assumption.
First, and most important, is that we can approximate the excitonic operator as a real bosonic operator,
thus neglecting the exciton structure.
Under such assumption the two-particle propagator $\Lmm^{<,eh}(t,t)$ becomes a one-particle propagator in the
excitonic representation.
We can then assume some effective exciton-exciton or exciton-phonon or exciton-photon interaction~\cite{Moskalenko2000} and
consider the corresponding dynamical self-energy.
The resulting KBE can then be written in terms of the retarded and advanced
$\Lmm^{(r/a)}$, which indeed have resonances at the excitonic poles and, together with the same approximations
used in the IP case (i.e. GKBA+Markov) we obtain:
\begin{equation}
\label{eq:KBE_pop_EXC_scatt_bound}
\partial_t \G^{(2)}_{x\blq,x\blq}(t) -iH^{ext}_{x\blq,x\blq}[\r(t)] =
-\sum_{\blq} \beta_{x\blq x\blq'}\big[ (1+N_{\blq'}) N_{\blq}+ N_{x\blq'} (1+N_{x\blq})\big] .
\end{equation}
The first difference compared to Eq.~(\ref{eq:KBE_pop_IP_scatt}) is of course that now the occupation factors
appear in the form $N(1+N)$, as opposite to the form $f(1-f)$ in the IP case, due to the different commutation
relation of bosonic and fermionic operators.
As a result, once the external potential is zero, a Bose distribution for $N_{x\blq}$ is a stationary solution of
Eq.~(\ref{eq:KBE_pop_EXC_scatt_bound}) while a Fermi distribution for $f_{n\blk}$ is solution of Eq.~(\ref{eq:KBE_pop_IP_scatt}).
We underline that a proper general derivation from first principles must take into account the
internal structure of the exciton and hence of the excitonic propagator.
This is however beyond the goals of the present manuscript.

Let us just comment more on Eq.~(\ref{eq:KBE_pop_EXC_scatt_bound}).
First of all it has the same structure of the equation typically used for exciton dynamics in the literature~\cite{Maialle1993}.
Due to the poles of $L^{(r/a)}$
the energy conservation will be of the form $E_{x\blq}-E_{x\blq'}=\w_{\l \blq-\blq'}$.
This is consistent with what can be obtained using the Fermi golden rule and assuming that the many-body
wave-function is well approximated using the excitonic wave-function
$|\Psi^{MB}_{\g\blq}\rangle \approx \excdop |\Psi_g \rangle$.
Such assumption was used for example in the description of the exciton-phonon scattering~\cite{MolinaSanchez2017} or
in describing the radiative recombination of excitons in Refs.~\cite{Palummo2015}.
Finally we remark that one would expect that, eventually,
the distribution could condensate if the temperature reached is low enough~\cite{Moskalenko2000}.
However an exciton condensate is not simply a non coherent distribution of 
excitons in the same level. It is a coherent state and thus, it must also have finite elements $\rho_\g$.
This is the well known off-diagonal long range order of a condensate.
On the contrary Eq.~\eqref{eq:KBE_pop_EXC_scatt_bound}
can only produce a finite temperature Bose distribution of population.

\section{Experimental measure of non-coherent excitons from Time--Resolved photoemission}\label{sec:detection}
%====================================================
We have discussed the equations of motion describing the formation of coherent and non-coherent excitons
and then speculated on their possible evolution due to the collision integral.
As a last step let us consider again the signature of non coherent excitons in photo-emission.
Here we follow our previous work of Ref.~\cite{Perfetto2016}. There we have discussed how to capture the signature of an exciton
in photo-emission, computing the Fourier transform of $G_{c\blk c\blk}^<(t,t')$.
For the case were a single exciton is present th exact $G^<(\w)$ can be computed analytically and has a pole at 
$E_{x\blq}-\e_{v\blk}$. The same result can be obtained within MBPT considering the T-matrix approximation to the
self-energy for $G^{(r/a)}(\w)$ and then computing $G^<(\w)$ using the following ansatz
\begin{equation}
G^{<}_{c\blk c\blk}(\w)=-f_c(\w)[G^{\rm (r)}_{c\blk c\blk}(\w)-G^{\rm (a)}_{c\blk c\blk}(\w)],
\label{eq:quasi_fluc_diss}
\end{equation}
which comes from the idea of an approximate fluctuation and dissipation theorem in the quasi-stationary case.
We underline here the similarity with the GKBA
\begin{equation}
G^{<}_{c\blk c\blk}(\w)=-f_{c\blk}[G^{\rm (r)}_{c\blk c\blk}(\w)-G^{\rm (a)}_{c\blk c\blk}(\w)],
\label{eq:GKBA}
\end{equation}
In Eq.~(\ref{eq:quasi_fluc_diss}) however if $G^{(r/a)}(\w)$ have more than one pole,
such poles are weighted differently by the energy dependent Fermi distribution
$f(\w)$. In the non correlated case, the density--matrix $\rho_{n\blk,n\blk}$ tends indeed
to a Fermi distribution $f(\e_{c\blk})$ and $G^{(r/a)}_{c\blk c\blk}(\w)$
have a single pole at $\w=\e_{c\blk}$. Thus both Eq.~(\ref{eq:quasi_fluc_diss}) and Eq.~~(\ref{eq:GKBA}) give
\begin{eqnarray}
\label{eq:G_lesser_IP}
G_{c\blk c\blk}^<(t,t')= f(\e_{c\blk})\, e^{-i\e_{c\blk}(t-t')}.
\end{eqnarray}
The approach with MBPT, together with Eq.~\eqref{eq:quasi_fluc_diss}, was then used for the case a finite excited density exist,
under the assumption of a quasi-stationary distribution of particles, i.e. a specific Fermi distribution
for $f_c(\w)$.
The idea of introducing a Fermi distribution is due to the fact that, to compute photo-emission
the one-particle $G^<$, i.e. a Fermion propagator, is used. However a Fermi distribution strictly
holds only for non interacting particles, i.e. it minimizes the IP free--energy $H=U-TS$
with $U$ the energy, $S$ the entropy and $T$ the temperature of the system.
Here, instead, we are dealing with excitons which, in the one particle picture,
are strongly correlated objects. Moreover excitons are (composite) Boson and, at least in the low density regime,
they are expected to distribute according to a Boltzmann~\cite{Rustagi2018,Steinhoff2017} or a Bose~\cite{Moskalenko2000} function.
A Bose function is also what we obtained in the previous section for the bound region of the spectrum.

The use of different distributions characterizes three recent papers~\cite{Perfetto2016,Rustagi2018,Steinhoff2017}
where the exciton signature in photoemission has been discussed.
The approaches agree on the main feature: nearby the QP peak $\e_{c\blk}$, a satellite at energy $E_{x\blq}-\e_{v\blk}$ is
observed.
However they differ in some aspects. The two works based on MBPT~\cite{Perfetto2016,Steinhoff2017} are very close.
They both use an almost identical statistical relation
involving a Fermi distribution as a starting point: a local relation (reported here in Eq.~\eqref{eq:quasi_fluc_diss})
in one case~\cite{Perfetto2016} and its integrated version (Eq.~(1) of Ref.~\cite{Semkat2009})
in the other~\cite{Steinhoff2017}. In the modellistic approach of Ref.~\cite{Rustagi2018} instead temperature
enters only via Boltzmann factors. As a result the dispersion of the excitonic pole is different.
One of course must consider that a modellistic approach is based on strong simplifications of the problem.
Here however we want to focus on our MBPT approach and consider how the approach could be changed. 

The first obvious change would be to replace the fermionic distribution for the QPs in the statistical relation
by a bosonic-like distribution for the exciton.
This can be done introducing the following generalization of Eq.~\eqref{eq:quasi_fluc_diss}:
\begin{eqnarray}
\label{eq:quasi_fluc_diss_ext}
G^{<}_{c\blk c\blk}(\w)&=&-W_c(\w)[G^{\rm (r)}_{c\blk c\blk}(\w)-G^{\rm (a)}_{c\blk c\blk}(\w)], \\
W_c(\w)&=&       b_x(\w) \theta(\e_{cbm}-\w)+f_{c}(\w)\theta(\w-\e_{cbm}),
\end{eqnarray}
where  $\e_{cbm}$ is the energy of the conduction band minimum. Thus the QP pole above $\e_{cbm}$ is weighted by
a Fermi distribution, while the correlated pole below by a Bose distribution $b(\w)$.
This sharp factorization in two regions of the spectrum reminds the chemical picture of Ref.~\cite{Semkat2009}. It
could be improved
using a smoother interpolation between the ``Bose like region'' and the ``Fermi like region'' in case the
distinction between bound and non bound states is not sharp.
However, in Eq.~\eqref{eq:quasi_fluc_diss_ext}, the thermal distribution
does not enter as a weighting term only, but also in the definition of the spectral function
$G^{\rm (r)}_{c\blk c\blk}(\w)-G^{\rm (a)}_{c\blk c\blk}(\w)$, since the latter must be
evaluated on thermal NEQ--correlated states to contain the excitonic pole. To this end a thermal
NEQ--IP states is used as a starting point.
Such initial states is chosen, again, with lowest IP free--energy, i.e. a Fermi distribution.
Correlations (the T--matrix in this case) are then switched on to reach the NEQ--correlated state. 
Here a fundamental question arise: while correlations are switched on,
is the NEQ--IP state with lowest IP free--energy connected to the NEQ correlated (excitonic) state with lowest correlated free--energy.
This is the equivalent of the equilibrium issue: is the EQ--IP state with minimum energy connected to the EQ correlated state at minimal
energy. The discussion of this question would further clarify the difference between MBPT and modellistic approaches. In particular on the
dispersion of the excitonic pole.

\section*{Conclusions \& Outlooks}

In the present manuscript we have proposed a set of two equations,
Eqs.~\eqref{eq:KBE_coh_EXC}-\eqref{eq:KBE_pop_EXC}, which can be implemented
in a first principles manner to describe the formation of coherent and non-coherent excitons.
In doing so we also highlighted how the resulting dynamics is related to the definition
of coherent states used in quantum optics. The implementation would
require the propagation of two vectors ($\r_{\g\blq}$ and the diagonal $\G_{\g\blq,\g\blq}$) whose size is
$N_{exc}\times N_{\blq}$ with $N_{exc}=n_c\times n_v\times n_\blk$
the number of excitonic bands, $N_{\blq}$ the number of q--points used to sample the BZ, $n_c$ and $n_v$ the number
of states in conduction and valence, and, finally, $n_\blk$ the number of k--points in the BZ.
As a comparison the ai-NEQ-MBPT approach, which we implemented and used to describe the generation of carriers,
propagates $n_\blk$ matrices of size $(n_c+n_v)\times(n_c+n_v)$.
Of course the size of the excitonic vector is bigger than the size of the carriers matrices,
since the phase space of two (correlated) particles is
bigger than the phase space of a single particle.
However, since we only need the diagonal of $\Gmm$ here, the problem is strongly simplified. 
It is equivalent to consider only the diagonal of the BSE matrix.
Moreover one can reasonably select few excitons resonant with the frequency of the external perturbation
and matching its q--point ($\blq=0$ for optical pulses).

We also discussed how the created non-coherent excitons would evolve
in time and thermalize with Eqs.~\eqref{eq:KBE_coh_EXC_scatt}-\eqref{eq:KBE_pop_EXC_scatt}.
The implementation would become more demanding in this case.
Indeed the carriers thermalization, under the approximations discussed, only involves $f_{n\blk}=\r_{n\blk,n\blk}$, i.e.
$\blq=0$ since $\blk=\blp$ (although $f_{n\blk}$ is coupled with $f_{n'\blk'}$). On the contrary
the exciton thermalization couples excitons with different transferred momentum. Nevertheless the implementation
would still be feasible, in particular if only the lowest energy excitonic bands are considered.
Indeed a similar approach has been implemented for excitons described by the Wannier equation, even considering the role of coherences
in the collision integral~\cite{Thranhardt2000,Selig2017,Berghauser2018}. In these works
the generation of excitons populations is due to the decay of coherent excitons via electron--phonon
interaction. In the present manuscript, instead, we have mostly considered their generation via the
interaction of coherent excitons with the external pulse, similarly to what done in the IP case.

Finally we considered a simple model, with a strongly bound excitonic band well separated from
the continuum and in the low pumping regime, to analyze the derived equations.
This is the same model we used  used in Ref.~\cite{Perfetto2016} to describe the signature
of non coherent excitons in photoemission spectra. Accordingly we also discussed this topic, comparing our
approach with recent works. We proposed, with Eq.~\eqref{eq:quasi_fluc_diss_ext}, a generalization of the
expression used to describe non-coherent exciton signature in photo-emission in Ref.~\cite{Perfetto2016}.
Moreover we critically discussed a possible issue of MBPT performed on top of NEQ states with
minimal free--energy, i.e. for NEQ thermal states.

\section*{Acknowledgements}
DS, AM and EP acknowledges the funding received from the European
Union project MaX Materials design at the eXascale H2020-
EINFRA-2015-1, Grant Agreement No. 676598 and Nano-
science Foundries and Fine Analysis - Europe H2020-INFRAIA-
2014-2015, Grant Agreement No. 654360.

\appendix

\section{Multiple excitons in the same quantum state}\label{App:bound_excitons}
%====================================================
We here build up the wave--function which includes an arbitrary number $N_x$ of bound excitons. 
$|\Psi_{N_x}\rangle$ can be obtained
as applying $N_x$--times the excitonic operator $\excdop$ to the ground state.

\begin{equation}
|\Psi_{N_x}\rangle=\prod_{\g\blq=1}^{N_{eh}}\sum_{cv\blk} A^{\g\blq}_{cv\blk} \hat{c}^\dag_{\blk+\blq}\hat{v}_{\blk} |\Psi_0\rangle
\end{equation}
In the IP case $A^{\g\blq}_{cv\blk}=\d_{c,c_0}\d_{v,v_0}\d_{k,k_0}$, then only one electron--hole pair can 
enter each $\{\g\blq\}$ state, i.e. the Fermi like character of electrons and holes is preserved.
The same holds in an approximate way for non-bound and weakly bound states in the correlated case,
where $A^{\g\blq}$ is very localized in $\blk$-space.
On the contrary for bound states $A^{\g\blq}_{cv\blk}\approx 1/N_{cv\blk}$ ($1/\callL$ in out model).
If the state $\{x\blq\}$ is occupied when the first photon is absorbed by the system the second
state can in general be any of the $\{\g\blq\}$ also including the state $\{x\blq\}$,
and the Bose like character of excitons emerges as explained below.

Let's consider for example the case with tho excitons. \\
The wave--function will be
\begin{eqnarray}
|\Psi_2\rangle
         &=& \hat{e}^\dag_{\g_2\blq_2} \hat{e}^\dag_{\g_1\blq_1} | \Psi_0 \rangle  \\
         &=&\sum_{c_2v_2\blk_2} A^{\g_2\blq_2}_{c_2v_2\blk_2} \hat{c_2}^\dag_{\blk_2+\blq_2}\hat{v_2}_{\blk_2}
            \sum_{c_1v_1\blk_1} A^{\g_1\blq_1}_{c_1v_1\blk_1} \hat{c_1}^\dag_{\blk_1+\blq_1}\hat{v_1}_{\blk_1} 
            | \Psi_0 \rangle 
\end{eqnarray}
Since ($\hat{c_2}^\dag_{\blk_2+\blq_2}\hat{v_2}_{\blk_2} \hat{c_1}^\dag_{\blk_1+\blq_1}\hat{v_1}_{\blk_1}) | \Psi_0 \rangle =0$
if 
\{${\hat{c_2}^\dag_{\blk_2+\blq_2}=\hat{c_1}^\dag_{\blk_1+\blq_1}}$ or
${\hat{v_2}_{\blk_2}=\hat{v_1}_{\blk_1}}$\}, $|\Psi_2\rangle$
can be re-written as
\begin{equation}
|\Psi_2^{MB}\rangle=\sum_{c_1v_1\blk_1} \sideset{}{'}\sum_{c_2v_2\blk_2}
           A^{\g_2\blq_2}_{c_2v_2\blk_2} \hat{c_2}^\dag_{\blk_2+\blq_2}\hat{v_2}_{\blk_2}
           A^{\g_1\blq_1}_{c_1v_1\blk_1} \hat{c_1}^\dag_{\blk_1+\blq_1}\hat{v_1}_{\blk_1} 
          |\Psi_0\rangle
\end{equation}
where the prime on the second sum means the terms $\{c_1,\blk_1+\blq_1\}=\{c_2,\blk_2+\blq_2\}$ and $\{v_1,\blk_1\}=\{v_2,\blk_2\}$
are excluded.
The second summation does not define exactly the creation of an excitonic state 
${\hat{e}^{\dag}_{\g_2\blq_2}}$ due to the ``prime'' in the summation. 
However, as long as $\hat{e}^{\dag}_{\g_2\blq_2}$ spans an infinite number of components
in the $\{cv\blk\}$ space, then
the prime in the summation can be neglected; it amounts in removing one point from the whole BZ,
thus in removing a null dimension set.
Only once a finite density of excitons is 
considered, the summation will start to differ from the definition of the exact excitonic operator,
since the dimension in the BZ will start to be different from zero.
How much this will deviate depends on the density of excitons (which defines the size of the zone
to be removed in the BZ) and the strength of the e--h interaction (i.e. on how much
$\{cv\blk\}$ space is spanned by the vector $A^{\g\blq}_{cv\blk}$).

\section{Coherent states}\label{App:coherent_states}
%====================================================
In the present manuscript we have considered the coherent state (Eq.~\eqref{eq:coherent-fock}), its fully coherent 
excitonic version (Eq.~\eqref{eq:fully_coher_exc}) and the partially coherent excitonic state 
(Eqs.~\ref{eq:part_coher_exc}). We inspect here their relation.
We start considering the fully coherent excitonic state at $t=0$.
It was constructed assuming the low pumping regime 
(or low density limit) and the delocalization of the excitonic wave-function in $\blk$ space.
Rewriting the sum as an exponent ($\sum_n=0^\infty x^n/n!=e^x$), using the definition of the
excitonic operator, and writing the sum of exponents as a product ($e^{\sum_\blk f_\blk}=\prod_\blk e^{f_\blk}$),
we obtain
\begin{equation}
\label{eq:fully_coher_exc_cvk}
|\Psi_{\infty}(\blq,0)\rangle = e^{\frac{-|\alpha|^2}{2}}\prod_\blk e^{\a A^{x\blq}_{\blk} \ecdop\evop }|\Psi_g\rangle
\end{equation}
If we drop the assumption on the excitonic wave-function discussed in app.~\ref{App:bound_excitons} and, 
on the contrary move to the situation where Pauli exclusion principle becomes dominant, we need to expand the exponent in
Eq.~\eqref{eq:fully_coher_exc_cvk}, thus obtaining the well known expression for the BCS ground state~\cite{Fetter-book}
\begin{equation}
|\Psi_{BCS}(\blq,0)\rangle= e^{\frac{-|\alpha|^2}{2}}\prod_\blk (1+\a A^{x\blq}_{\blk} \ecdop\evop )|\Psi_g\rangle
\end{equation}
Finally linearizing the latter to single excitations we end up with the expression
\begin{equation}
|\Psi_{eh}(\blq,0)\rangle= e^{\frac{-|\alpha|^2}{2}}\sum_\blk (1+\a A^{x\blq}_{\blk} \ecdop\evop )|\Psi_g\rangle
\end{equation}
which gives Eqs.~\ref{eq:part_coher_exc}.
In the IP case we just considered the counterpart of this latter (\ref{eq:part_coher_ip}).
Of course there is no IP equivalent of Eq.~\eqref{eq:fully_coher_exc_cvk}, since in the IP case the
excitonic eigen-vector $A^{\g\blq}_{\blk}$ reduces to a delta which is localized in k-space by definition.
We did not consider the BCS like state for the exciton and its IP counterpart, since we are not 
discussing here the high density regime.

\bibliographystyle{unsrt}
\bibliography{manuscript}

\end{document}